\documentclass [12pt]{article}

\usepackage{natbib}
\usepackage {graphicx}
\usepackage{epsfig,amsmath,latexsym,amssymb}
\usepackage{bm}

\usepackage{booktabs}
\usepackage{multirow}
\begin{document}


\title{\vspace{-2cm} Fast Numerical Method for Pricing of Variable Annuities with Guaranteed Minimum Withdrawal Benefit under Optimal Withdrawal Strategy}

\author{Xiaolin Luo$^{1,\ast}$ and Pavel V.~Shevchenko$^{2}$}

\date{\footnotesize{Draft, this version 24 October 2014 }}

\maketitle

\begin{center}
\footnotesize { \textit{$^{1}$ The Commonwealth Scientific and Industrial Research Organisation, Australia; e-mail: Xiaolin.Luo@csiro.au \\
$^{2}$ The Commonwealth Scientific and Industrial Research Organisation, Australia;
e-mail: Pavel.Shevchenko@csiro.au  \\
$^*$ Corresponding author} }
\end{center}

\begin{abstract}
\noindent A variable annuity contract  with Guaranteed Minimum
Withdrawal Benefit (GMWB) promises to return the entire initial
investment through cash withdrawals during the policy life plus the
remaining account balance at maturity,  regardless of the portfolio
performance. Under the optimal withdrawal strategy of a policyholder, the pricing of variable  annuities with GMWB becomes an
optimal stochastic control problem.   So far in the literature these
contracts have only been evaluated by solving partial differential
equations (PDE) using the finite difference method. The well-known
Least-Squares or similar Monte Carlo methods cannot be applied to
pricing these contracts because the paths of the underlying wealth
process are affected by optimal cash withdrawals (control variables)
and thus cannot be simulated forward in time.
  In this
paper we present a very efficient new  algorithm for pricing these
contracts  in the case when transition density of the underlying
asset between withdrawal dates or its moments are known.  This
algorithm  relies on computing the expected contract value through a
high order Gauss-Hermite
 quadrature applied on a cubic spline interpolation. Numerical
results from the new algorithm for  a series of GMWB contract are
then presented, in comparison with  results using the finite
difference method solving corresponding PDE. The comparison demonstrates that the
new algorithm produces  results in very close agreement with those
of the finite difference method, but at the same time it is
significantly faster; virtually instant results on a standard desktop PC.

\vspace{0.2cm} \noindent \textbf{Keywords:} {Variable Annuity,
Optimal Stochastic Control, Guaranteed Minimum Withdrawal Benefit,
Gauss-Hermite Quadrature,  Cubic Spline.}
\end{abstract}

\pagebreak

\section{Introduction}
\label{sec:introduction} The world population is becoming older
fast. According to world population data recently published in \citet{UN2013} there were about 800 million persons aged 60
years or over in 2012, but this number is expected to grow to 2
billion by 2050. The oldest old (aged 80 or over) population
accounted for 14\% of the older population (aged over 60) and this
is expected to grow to 20\% by 2050. The number of of centenarians
(aged 100 years or over) is growing even faster.
  It is projected to increase tenfold, from approximately
343,000 in 2012 to 3.2 million by 2050. The old-age support ratio
(number of persons aged 15 to 64 years per person aged 65 years or
over) is 8 to 1 now, but it will be reduced to 4 to 1 by 2050. As a
consequence the age-related spending is projected to rise
dramatically in the coming decades in all the developed countries.
Increasingly governments in the developed world realize they cannot
afford paying sufficient public pensions and are looking for
innovations in the financial market for providing
 some urgently needed solutions. Variable annuity is one of the
 products that can help with meeting the challenges posed by the so
 called longevity risk.

 In this paper we consider a variable annuity contract with
Guaranteed Minimum Withdrawal Benefit (GMWB) that promises to return the
entire initial investment through cash withdrawals during the policy
life plus the remaining account balance at maturity, regardless of
the portfolio performance. Thus even  when the account of the
policyholder falls to zero before maturity, GMWB feature will
continue to provide the guaranteed cashflows. GMWB allows the policyholder to withdraw funds below or at contractual rate without
penalty and above the contractual rate with some penalty. If the
policyholder behaves passively and the withdraw amount at each
withdrawal date
 is predetermined at
the beginning of the contract, then the behavior of the
 policyholder is  called ``static''. In this case the paths of the account  can be simulated  and a standard Monte Carlo
 simulation method can be used to price the GMWB. On the other hand if the policyholder optimally decide the amount of withdraw
 at each withdrawal date, then   the behavior of the policyholder is  called ``dynamic''.
A rational policyholder of GMWB will always choose the optimal
withdrawal strategy to maximize the present value of cash flows
generated from holding the GMWB. Under the optimal withdrawal
strategy of a policyholder, the pricing of variable annuities with
GMWB becomes an optimal stochastic control problem. This problem
cannot be solved by a simulation based method such as the well known
Least-Squares Monte Carlo method introduced in \citet{Longstaff2001}, due to the fact that the paths of
the underlying wealth process are altered by the optimal cash withdrawals that should be found from backward solution and the underlying wealth process cannot be
simulated forward in time.

The variable annuities with GMWB feature have been considered in a
number of papers over the last decade.
  \citet{milevsky2006financial} developed a variety of methods for
  pricing GMWB products. In their {\it static} approach the GMWB product
  is decomposed into a Quanto Asian put plus a generic term-certain
  annuity. In their {\it dynamic} approach they assume the
  policyholder can terminate (surrender) the contract at the optimal time, which
  leads to an optimal stopping problem akin to pricing an American
  put option. \citet{bauer2008universal}
 presents valuation of variable annuities
with multiple guarantees. In their {\it dynamic} approach a strategy
 not only consists of the decision whether or not to surrender, but
also a numerous possible withdrawal amounts at each payment date.
They have developed a multidimensional discretization approach in
 which  the Black-Scholes PDE is  transformed to a
one-dimensional heat equation and a quasi-analytic solution is
obtained through a simple piecewise summation  with a  linear
interpolation on a mesh. Unfortunately the numerical formulation
considered in \citet{bauer2008universal} has four dimensions and the
computation of even a single  price of the GMWB contract under the
optimal policyholder strategy is very costly; it is mentioned in their paper that it took more than 15
hours CPU on an Intel Pentium IV with 2.80 GHz and 1 GB RAM, and no
results for the dynamic case were shown. \citet{dai2008guaranteed} developed
an efficient finite difference algorithm using the penalty
approximation to solve the singular stochastic control problem for a
continuous withdrawal model under the dynamic (optimal) withdrawal
strategy.
 They have also developed a finite difference algorithm for the
  more realistic discrete withdrawal formulation.
 Their results show that the GMWB values from the
 discrete model converge to those of the continuous model.  \citet{Forsyth2008}
present an impulse stochastic control formulation for pricing
variable annuities with GMWB under the optimal policyholder
behavior, and develop a single numerical scheme for solving the
Hamilton-Jacobi-Bellman variational inequality for the continuous
withdrawal model  as well as for pricing the discrete withdrawal
contracts.
 In \citet{bacinello2011unifying}
  the {\it static} valuations are performed via ordinary Monte Carlo method, and the  {\it mixed}
  valuations, where the policyholder is `semiactive' and can decide to  surrender the
 contract at any time during the life of the GMWB contract, are
 performed by the Least-Squares Monte Carlo method.

 In the case when transition density of the underlying wealth process between withdrawal dates or its moments are known in closed form, often it can be more convenient and more efficient to utilize direct integration methods to calculate the required option price expectations in backward time-stepping procedure. In this paper we present a method that relies on computing the expected option values in a backward time-stepping between
 withdrawal dates through Gauss-Hermite integration quadrature applied on a cubic spline interpolation. For convenience, hereafter  we
 refer this new algorithm as GHQC (Gauss-Hermite quadrature on cubic spline). We adopt the method developed in \citet{LuoShevchenkoGHQC2014} for pricing American options and extend it to solving optimal stochastic control problem for pricing GMWB variable annuity. This allows to get virtually instant results for typical GMWB annuity prices on the standard desktop PC. In this
paper we consider pricing
 variable  annuities with GMWB under both static and dynamic (optimal) policyholder behaviors.
  Here our definition of
 ``dynamic" is similar to the one used by \citet{bauer2008universal},
   \citet{dai2008guaranteed} and \citet{Forsyth2008}, i.e. the rational policyholder can decide an
   optimal amount to withdraw at each discrete payment date to
   maximize the expected discounted value of the cash flows
   generated from holding the variable annuity with GMWB.
  At discrete withdrawing
dates the proper jump conditions are applied at various withdrawal
account levels,   allowing the optimal withdrawal decision to be
made by choosing a withdrawal amount to maximize the cashflow. Once
the fair price of GMWB for given inputs is evaluated, the fair
fee charged by the policy issuer can be determined
 iteratively by matching the initial premium and the fair price.

In the next section we describe the GMWB product with discrete
withdrawals, the underlying stochastic model and the optimization
problem. Section \ref{GHQCalgo_sec} presents the GHQC algorithm for
pricing the GMWB
 contracts under both static and dynamic (optimal) policyholder behaviors.
 In Section \ref{NumericalResults_sec},  numerical results  for the fair fees
 under a series GMWB contract conditions are presented, in comparison with the results from solving PDEs using finite
 difference method. The comparison   demonstrates that the new algorithm produces  results very close to those of the finite difference method,
  but at the same time  it is significantly faster. Concluding remarks are given in Section \ref{conclusion_sec}.

\section{Model}\label{model_sec}
 Consider the  GMWB annuity contract and underlying asset stochastic model as follows.

\begin{itemize}
\item Let $S(t)$ denote the value of the reference portfolio of assets
(mutual fund index, etc.) underlying the variable annuity policy at
time $t$ that under no-arbitrage condition follows the risk neutral stochastic process
\begin{equation}\label{referenceportfolio_eq}
dS(t)=r(t) S(t) dt+\sigma(t) S(t) dB(t),
\end{equation}
where $B(t)$ is the standard Wiener process, $r(t)$ is risk free
interest rate and $\sigma(t)$ is volatility. For simplicity hereafter we assume that model parameters are piece-wise constant functions of time for time discretization $0=t_0<t_1<\cdots<t_N=T$, where $t_0=0$ is today and $T$ is annuity contract maturity. Denote corresponding asset values as $S(t_0),S(t_1),\ldots,S(t_N)$; and risk-free interest rate and volatility as $r_1,\ldots,r_N$ and $\sigma_1,\ldots,\sigma_N$ respectively. That is $r_1$ is the interest rate for time teriod $(t_0,t_1]$; $r_2$ is for $(t_1;t_2]$, etc and similar for volatility.

\item The premium paid by policyholder upfront at $t_0$ is invested into the reference portfolio of risky assets $S(t)$.
Denote the value of this variable annuity account (hereafter
referred to as \emph{wealth account}) at time $t$ as $W(t)$, i.e.
the upfront premium paid by policyholder is $W(0)$. GMWB guarantees
the return of the premium via withdrawals $\gamma_n\ge 0$ allowed at
times $t_n$, $n=1,2,\ldots,N$.  Let $N_w$ denote the number of
withdrawals in a year (e.g. $N_w=12$ for a monthly withdrawal), then
the total number of withdrawals $N=\lceil\; N_w\times T \;\rceil$,
where $N=\lceil\; \cdot\;\rceil$ denotes  the ceiling of a float
number. The total of withdrawals cannot exceed the guarantee $W(0)$
and withdrawals can be different from contractual (guaranteed)
withdrawal $G_n=W(0)(t_n-t_{n-1})/T$,  with penalties imposed if
$\gamma_n>G_n$. Denote the annual contractual rate as $g=1/T$.

\item Denote the value of the guarantee at time $t$ as $A(t)$, hereafter referred to as \emph{guarantee account}. Obviously, $A(0)=W(0)$. For clarity of notation, denote the time immediately before $t$ as  $t^-$, and  immediately
after $t$ as  $t^+$, and functions of time are left continuous, e.g. $A(t_n)=A(t_n^+)$ etc. Then the guarantee balance
evolves as
\begin{equation}\label{accountbalance_eq}
A(t_n^+)=A(t_n^{-})-\gamma_n=A(t^+_{n-1})-\gamma_n,\;\; n=1,2,\ldots,N
\end{equation}
with $A(T^+)=0$, i.e. $W(0)=A(0) \ge \gamma_1+\cdots+\gamma_N$ and
$A(t_{n-1})=A(t_{n-1}^{+})\ge \sum_{k=n}^N\gamma_{k}$. The account balance $A(t)$ remains
unchanged within the interval $(t_{n-1},\;t_n), \;n=1,2,\ldots,N$.
\item In the case of reference portfolio process (\ref{referenceportfolio_eq}), the wealth account $W(t)$ evolves as
\begin{equation}
W(t_n)=\max\left[W(t_{n-1})
e^{(r_n-\alpha-\frac{1}{2}\sigma_n^2)dt_n+\sigma_n \sqrt{dt_n} z_n} -
\gamma_n,0\right],\;\; n=1,2,\ldots,N,
\end{equation}
where $dt_n=t_n-t_{n-1}$, $z_n$ are iid standard Normal random
variables and $\alpha$ is the annual fee charged by the insurance company. If the account balance
becomes zero or negative, then it will stay zero till maturity. Then
from $t=t_{n-1}^+$ to $t=t_n^-$ the  value of wealth account $W(t)$ evolves as
\begin{equation}\label{eq_Wt}
W(t_n^-)=W(t_{n-1}^+)
e^{(r_n-\alpha-\frac{1}{2}\sigma_n^2)dt_n+\sigma_n \sqrt{dt_n}
z_n},\; \;n=1,2,\ldots,N.
\end{equation}
Note the process for $W(t)$ within $(t_{n-1},t_n)$ differs from the process for the
underlying asset $S(t)$ only in the drift; the former has a reduced
drift due to the continuously charged fee $\alpha$ on the GMWB contract.

\item The cashflow received by the policyholder at withdrawal time $t_n$ is given by
\begin{equation}
C(\gamma_n)=\left\{\begin{array}{ll}
                   \gamma_n, & \mbox{if}\; 0\le \gamma_n\le G_n, \\
                   G_n+(1-\beta)(\gamma_n-G_n), & \mbox{if}\; \gamma_n>G_n,
                 \end{array} \right.
\end{equation}
where $G_n$ is contractual  withdrawal. That is, penalty is applied if
withdrawal $\gamma_n$ exceeds $G_n$, i.e. $\beta\in
[0,1]$ is the penalty applied to the portion of withdrawal above
$G_n$.

\item Denote the value of variable annuity at time $t$ as $Q_t(W(t),A(t))$, i.e. it is determined by values of the
wealth and guarantee accounts $W(t)$ and $A(t)$.  At maturity, the policyholder takes the maximum between
the remaining guarantee withdrawal net
of penalty charge and the remaining balance of the personal account,
i.e. the final payoff is
\begin{equation}
Q_{t_N^-}(W(T^-),A(T^-))=\max\left(W(T^-),C(A(T^-))\right).
\end{equation}
The policyholder receives cashflows $C(\gamma_n),\;\;
n=1,2,\ldots,N-1$ and the final payoff at maturity. The present
value of total payoff
 is
\begin{equation}
P_0=e^{-r_{0,N}T}\max\left(W(T^-),C(A(T^-))\right)+\sum_{n=1}^{N-1}
e^{-r_{0,n}t_n}C(\gamma_n),
\end{equation}
where $r_{i,n}=\frac{1}{t_n-t_i}\int_{t_i}^{t_n} r(\tau)d\tau$,
$t_n>t_i$.
\end{itemize}

Under the above assumptions/conditions, the fair no-arbitrage value of the annuity at time $t_n$
is
\begin{eqnarray}
&&\hspace{-1cm}Q_{t_n}\left ( W(t_n),A(t_n)\right)\nonumber\\
&&\hspace{-1cm}=\max_{\gamma_{n+1},\ldots,\gamma_{N-1}}\mathrm{E}_{t_n}\left[e^{-r_{n,N}(T-t_n)}\max\left(W(T^-),C(A(T^-))\right)+\sum_{j=n+1}^{N-1}
e^{-r_{n,j}(t_j-t_n)}C(\gamma_j)\right],
\end{eqnarray}
and the today's value of the annuity policy corresponds to $Q_0(W(0),A(0))$ which is a function of policy fee
 $\alpha$. 
Here, $\gamma_{1},\ldots,\gamma_{N-1}$ are the control variables chosen to maximize the expected value of discounted cashflows, and expectation $\mathrm{E}_t[\cdot]$ is taken under the risk-neutral process conditional on $W_t$ and $A_t$. The fair fee $\alpha=\alpha^\ast$ corresponds to $Q_0\left(W(0),A(0)\right)=W(0)$. It is important to note that control variables can be different for different realizations of underlying process and moreover the control variable $\gamma_n$ affects the transition law of the underlying wealth process from $t_n$ to $t_{n+1}$, i.e. calculating GMWB annuity price is solving optimal stochastic control problem.


\section{Numerical valuation of GMWB}\label{GHQCalgo_sec}
In the case of continuous withdrawal, following the procedure of
deriving the Hamilton-Jacobi-Bellman (HJB) equations in stochastic
control problems, the value of the annuity under optimal withdrawal
is found to be governed by a two-dimensional PDE; see \citet{milevsky2006financial}, \citet{dai2008guaranteed} and \citet{Forsyth2008}. For
discrete withdrawals, the governing PDE in the period between
withdrawing dates is one-dimensional, similar to the Black-Scholes
equation, with jump conditions at each withdrawing date to link the
prices at the adjacent periods. Below we consider an alternative
approach without dealing with PDEs.


The annuity price at any time $t$ for a fixed $A(t)$  is a
function of $W$ only. Note $ A(t_{n-1}^+)= A(t_{n}^-)=A$ is constant in the period
$(t_{n-1}^+,t_n^-)$. Thus in a backward time-stepping setting (similar to a
finite difference scheme) the option price at
 time $t=t_{n-1}^+$  can be evaluated as the following expectation
\begin{equation}\label{eq_expS}
Q_{t^+_{n-1}}\left(W(t_{n-1}^+), A\right)=\mathrm{E}_{t_{n-1}}\left[e^{-r_n dt_n}
Q_{t_n^{-}}\left(W(t_n^-),A\right)|W(t_{n-1}^+),A\right].
\end{equation}

Assuming the conditional probability
distribution density of $W(t_n^-)$ given $W(t_{n-1}^+)$  is known as
$p_n(w(t_n)|w(t_{n-1}))$, then the above expectation can be evaluated
by

\begin{equation}\label{eq_intS}
Q_{t_{n-1}^+}\left(W(t_{n-1}^+), A\right)=\int_0^{+\infty}
e^{-r_n dt_n} p_n(w|W(t_{n-1}^+)) Q_{t_n^-}(w,A)dw.
\end{equation}
In the case of wealth process (\ref{eq_Wt}) the transition density $p_n(w(t_n)|w(t_{n-1}))$ is known in closed form and we will use Gauss-Hermite quadrature for the evaluation of the
above integration over an infinite domain. The required continuous
function $Q_t(W,A)$ will be approximated by a cubic spline
interpolation on a discretized grid in the  $W$ space.

Any change of $A(t)$ only occurs at withdrawal dates. After the
amount $\gamma_n$ is drawn at $t_n$, the wealth account reduces
from $W(t_n^-)$ to $W(t^+_n) = \max (W(t_n^-) -\gamma_n,0)$, and the
guarantee balance drops from $A(t_n^-)$ to $A(t_n^+)=A(t_n^-) -
\gamma_n$.
 Thus the jump condition of $Q_t(W,A)$ across $t_{n}$ is given
by

\begin{equation}\label{eqn_jump}
Q_{t_{n}^-}(W(t_{n}^-),A(t_{n}^-))=\max_{0 \leq \gamma_n\leq A(t_{n}^-) } [Q_{t_n^+}(\max(W(t_{n}^-)-\gamma_n,0),
A(t_{n}^-)-\gamma_n)+C(\gamma_n)].
\end{equation}
For optimal strategy, we chose a value for $\gamma_n$ under the
restriction $0 \leq \gamma_n\leq A(t_n^-) $ to maximize the function value
$Q_{t_n^-}(W,A)$ in (\ref{eqn_jump}). Repeatedly applying (\ref{eq_intS}) and (\ref{eqn_jump}) backwards in time starting from
\begin{equation}
Q_{t^-_N}(W(T^-),A(T^-))=\max\left(W(T^-),C(A(T^-))\right)
\end{equation}
gives us annuity value at $t=0$.

\subsection{Overall algorithm description}

For a fixed guarantee balance $A$ between given withdrawal dates,
the price $Q_t(W,A)$  can be numerically evaluated  using
(\ref{eq_intS}).  For now we leave details of computing
(\ref{eq_intS}) to the next section and assume it can be done with
sufficient accuracy and efficiency. Starting from a final condition
at $t=T^-$ (just immediately before the final withdrawal), a
backward time stepping using (\ref{eq_intS}) gives solution up to
time $t=t_{N-1}^+$. In order to
 apply the jump condition at each withdrawal date and find the solution $Q_0(W=W(0),A=W(0))$,
 this backward time stepping needs to be done for many different
levels of $A$.  Applying jump condition (\ref{eqn_jump}) to the
solution at $t=t_{N-1}^+$ we obtain the solution at $t=t_{N-1}^-$
from which further backward time stepping gives us solution at
$t=t_{N-2}^+$, and so on. The numerical algorithm takes the
following key steps:

\begin{itemize}
\item Step 1. Generate an auxiliary finite grid  $0 = A_1 < A_2 < A_3
\cdots < A_J = W(0)$ to track the guarantee  account $A$.
\item Step 2.  Discretize wealth account $W$ space as $W_0
,W_1, \ldots,W_M$.
\item Step 3. At $t=t_N=T$ apply the final condition at each node point $(W_m, A_j)$, $j=1,2,\ldots, J$, $m=1,2,\ldots, M$
to get payoff $Q_{T^-}(W,C(A))$.
\item Step 4. Evaluate integration (\ref{eq_intS}) for each of the $A_j$  to obtain $Q_{t_{N-1}^+}(W, A)$.
\item Step 5. Apply the jump condition (\ref{eqn_jump}) to obtain $Q_{t_{N-1}^-}(W, A)$ for all possible jumps and find the jump to maximize
$Q_{t_{N-1}^-}(W, A)$.
\item Step 6. Repeat Step 4 and 5 for $t=t_{N-2}, t_{N-3}, \ldots, t_1$.
\item Step 7. Evaluate integration (\ref{eq_intS}) for the backward time step from $t_1$ to $t_0$ for the single node value
$A=A_J=W(0)$ to obtain solution  $Q_0(W, A_J)$ and take the value
$Q_0(A_J, A_J)$ as the annuity price at $t=t_0$.
\end{itemize}

Below we discuss details of the algorithm of  the numerical
integration of (\ref{eq_intS}) using  Gauss-Hermite quadrature on a
cubic spline interpolation, followed by the application of jump
conditions.

\subsection{Numerical evaluation of the expectation}\label{sec_GHQC}
Similar to a finite difference scheme, we propose to discretize the
asset domain $[W_{\min}, W_{\max}] $  by $W_{\min} =W_0 < W_1,
\ldots,W_M=W_{\max}$ , where $W_{\min}$ and $W_{\max}$ are the lower
and upper boundary, respectively.  For pricing GMWB, because of the
finite reduction of $W$ at each withdrawal date, we have to consider
the possibility of   $W$ goes to zero, thus the lower bound
$W_{\min}=0$. The upper bound is set  sufficiently
 far from the
spot asset value at time zero $W(0)$. A good choice of such a
boundary could be $W_{\max}=W(0) \exp(5\sigma \sqrt{T})$.  The idea
is to find option values at all these grid points at  each time step
from  $t_n^-$ to $t_{n-1}^+$  through integration  (\ref{eq_intS}),
starting at maturity $t=t_N^-=T^-$.  At each time step we evaluate
the integration (\ref{eq_intS}) for every grid point  by a high
accuracy numerical quadrature.

At time step $t_n^- \rightarrow t_{n-1}^+$, the option value at
$t=t_n^-$ is known only at grid points $W_m$, $m=0,1,\ldots,M$. In
order to approximate the continuous function $Q_t(W,A)$ from the
values at the discrete grid points, we propose to
 use the cubic spline interpolation
 which is smooth in the first derivative and continuous in the second derivative.
 The error of cubic
spline is $O(h^4)$ where $h$ is the size for the spacing of the
interpolating variable, assuming a uniform spacing. The cubic spline
interpolation involves solving a tri-diagonal system of linear
equations for the second derivatives at all grid points. For a fixed
grid and constant $r$ and $\sigma$,  the tri-diagonal matrix can be
inverted once and at each time
 step only the back-substitution in the cubic spline procedure is required.

The process for $W(t)$  between withdrawal dates is a standard stock
market process given in (\ref{eq_Wt}), the conditional density of
$W(t_n^-)$ given $W(t_{n-1}^+)$  is from a lognormal distribution,
as is evident from (\ref{eq_Wt}). A more convenient and common
practice is to work with  $\ln(W)$ which is from
normal distribution.  Note that when we use  $\ln(W)$, the
minimum $W_{\min}$ cannot
 be zero, and instead we have to set  $W_{\min}$ to be a very small value (e.g. $W_{\min}=10^{-10}$).

In order to make use of the highly efficient Gauss-Hermite numerical
quadrature for integration over an infinite domain, we introduce a
new variable
   \begin{equation}\label{eq_y}
Y(t_n)=\frac{\ln\left(W(t_n^-)/W(t_{n-1}^+)\right)-\nu_n}{\tau_n},
 \end{equation}
where $\nu_n=(r_n-\alpha-\frac{1}{2}\sigma_n^2 )dt_n$ and $\tau_n=\sigma_n\sqrt{dt_n}$ and denote the annuity price function $Q_t(w,\cdot)$ after this transformation as $Q_t^{(y)}(y,\cdot)$.
Apparently from (\ref{eq_Wt}), $Y(t_n)$ is from
standard normal distribution, thus by changing variable from $W(t_n)$ to
$Y(t_n)$ the integration (\ref{eq_intS}) becomes

 \begin{equation}\label{eq_intY}
Q_{t_{n-1}^+}\left(W(t_{n-1}^+), A\right)=\frac{
e^{-r_n dt_n}}{\sqrt{2\pi}} \int_{-\infty}^{+\infty}
e^{-\frac{1}{2}y^2} Q^{(y)}_{t_n^-}(y,A) dy,
\end{equation}
For an arbitrary function $f(x)$, the Gauss-Hermite
quadrature is
\begin{equation}\label{eq_GHQ}
\int_{-\infty}^{+\infty}e^{-x^2}f(x)dx \approx \sum_{i=1}^q
\lambda_i^{(q)} f(\xi_i^{(q)}),
\end{equation}
where  $q$ is the order of the Hermite polynomial, $\xi_i^{(q)}$ are
the roots of the Hermite polynomial $H_q(x) (i = 1,2,\ldots,q)$, and
the associated weights $ \lambda_i^{(q)}$  are given by

$$\lambda_i^{(q)}= \frac {2^{q-1} q! \sqrt{\pi}} {q^2[H_{q-1}(\xi_i^{(q)})]^2}.$$
In general, the abscissas and the weights  for the Gauss-Hermite quadrature for a given order $q$ can be readily computed, e.g. using functions in 
\citet{Pres92}; also as a reference, the abscissas and the weights for $q=5, 6, 16$ are presented in \citet{LuoShevchenkoGHQC2014}.

Applying a change of variable $x=y/\sqrt{2}$ and use the
Gauss-Hermite quadrature to (\ref{eq_intY}), we obtain
  \begin{eqnarray}\label{eq_qy}
Q_{t_{n-1}^+}\left(W(t_{n-1}^+), A)\right)&=&\frac{
e^{-r_n dt_n}}{\sqrt{\pi}} \int_{-\infty}^{+\infty} e^{-x^2} Q_{t_n^-}^{(y)}
(\sqrt{2}x,A) dx \nonumber\\
&\approx&  \frac{ e^{-r_n dt_n}}{\sqrt{\pi}} \sum_{i=1}^q
\lambda_i^{(q)} Q_{t_n^-}^{(y)}(\sqrt{2}\xi_i^{(q)},A).
\end{eqnarray}
If we apply the change of variable (\ref{eq_y}) and the
Gauss-Hermite quadrature (\ref{eq_qy}) to every grid point $W_m$,
$m=0,1,\ldots,M$, i.e. let $W(t_{i-1}^+)=W_m$, then the option values
at time $t=t_{i-1}^+$  for all the grid points  can be evaluated
through (\ref{eq_qy}).

As is commonly practiced in a finite difference setting for option
pricing, the working domain in asset space is in
 terms of $X=\ln (W/W(0))$, where $W(0)$ is the spot value at time $t=0$. In our implementation we have
 $X_{\min}=\ln(W_{\min}/W(0))$ and set $X_{\max}=\ln(W_{\max}/W(0))=5\sigma \sqrt{T}$. The domain $(X_{\min}, X_{\max})$ is
 uniformly discretised to yield the grid $(X_{\min}=X_0 , X_1=\delta X, X_2=2\delta X,\ldots, X_M=M\delta X=X_{\max})$,
  where $\delta X=(X_{\max}-X_{\min})/M$. The grid point $W_m$, $m=0,1,2,\ldots,M$, is then given by $W_m=W(0)\exp(X_m)$.

\begin{figure}[!h]
\begin{center}
\includegraphics[scale=0.5]{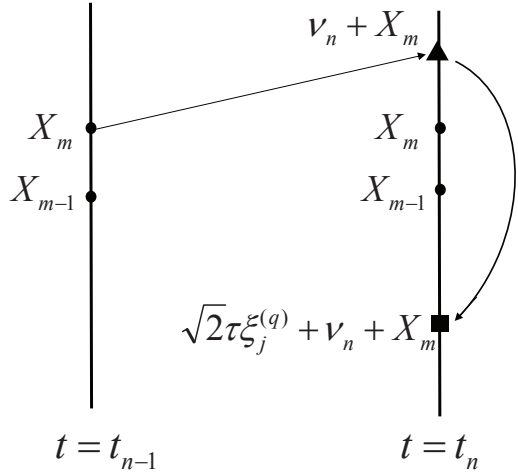}
\vspace{0cm} \caption{Illustration of Gauss-Hermite quadrature
application for an arbitrary grid point $X_m$ at time $t=t_{n-1}$.
The solid circles are fixed grid points, the solid triangle is the
 point of the expected mean at $t=t_n$ given $X_m$ at $t=t_{n-1}$,
 and the solid square is the $j-th$ quadrature point corresponding to $X_m$.} \label{fig_drawing}
\end{center}
\end{figure}

For each grid point $W_m$ or $X_m$, the variable $Y(t_n)$ is given by
(\ref{eq_y}) with $W(t_{i-1}^+)=W_m$, and the relationship between
$X(t_n)=\ln (W(t_n)/W(0))$ and $Y(t_n)$ for  $W_m$ is worked out to be $X(t_n)=\tau_n Y(t_n)
+\nu_n+X_m$, thus the numerical integration value for grid point $X_m$
at time $t_{n-1}^+$ can be expressed, from (\ref{eq_qy}), as

 \begin{equation}\label{eq_qX}
Q^{(x)}_{ t_{n-1}^+}\left(W(t_{n-1}^+), A\right) \approx  \frac{
e^{-r_n dt_i}}{\sqrt{\pi}} \sum_{i=1}^q \lambda_i^{(q)}
Q^{(x)}_{t_n^-}(\sqrt{2}\tau \xi_i^{(q)}+\nu_n+X_m, A),
\end{equation}
where $Q^{(x)}_{t_{n}}({X(t_n),A})$ denotes the option value as a
function of $X(t_n)=\ln(W(t_n)/W(0))$.
The above description of the
numerical integration using Gauss-Hermite quadrature is illustrated
in Figure \ref{fig_drawing}.

The continuous function $Q^{(x)}_{t_n}(x, A)$ is approximated by the
cubic spline interpolation based on variable $X$, given the values
$Q^{(x)}_{t_n}(X_m, A)$ at discrete points $X_m$,
$m=0,1,2,\ldots,M$.
 The cubic spline interpolation has a
much higher order of accuracy than linear or quadratic
interpolation. Natural boundary conditions are imposed at the two
ends $X_0 =X_{\min}$ and $X_M = X_{\max} $, i.e. we assume zero
second derivative of the spline function at the two ends.

\begin{figure}[!h]
\begin{center}
\includegraphics[scale=0.7]{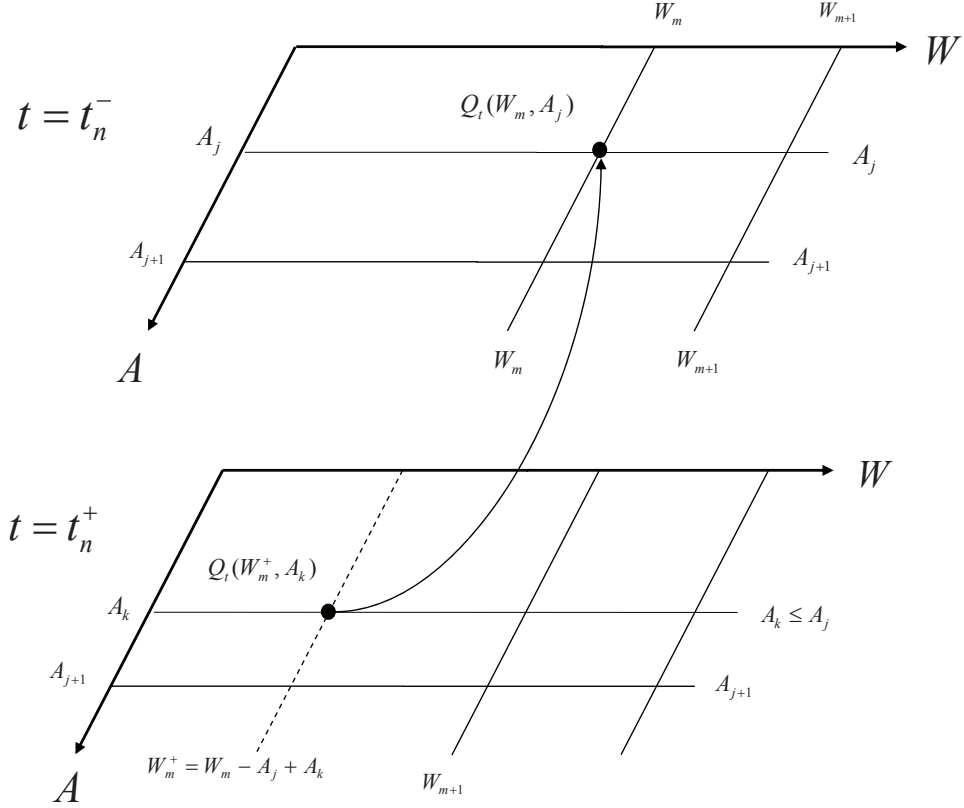}
 \caption{Illustration of jump conditions applied to finite
difference grids.} \label{fig1}
\end{center}
\end{figure}

\subsection{Jump condition application}
Let us introduce an auxiliary finite grid $0 = A_1 < A_2 < A_3
\cdots < A_J = W(0)$ to track the remaining guarantee balance  $A$,
where $J$ is the total number of nodes in the guarantee balance
amount coordinate. The upper limit $W(0)$ is needed because the
remaining guarantee balance cannot exceed the target initial account
value $W(0)$. For each $A_j $, we associate a continuous solution
from (\ref{eq_qX}) and the cubic spline interpolation.
 At every jump we let $A$ to be one of
the grid points $A_j ,\;1 \le j \le J$. Since $A$ is always known at
each jump to be one of the fixed nodal point values, there is no
need to continuously track the actual evolution of the guarantee
balance amount $A$ during the entire finite difference solving
process. In addition, we also limit the number of possible discrete
withdrawal amounts to be finite. The natural simple choice (though
not necessary) is to only allow the guarantee balance to be equal to
one of the grid points $0 = A_1 < A_2 < A_3 \cdots < A_J = W(0)$.
This implies that, for a given balance $A_j$ at time $t_n^-$, the
possible value after the withdraw at $t_n^+$ has to be one of the
grid points equal to or less than $A_j$, i.e.  $A_j^+=A_k$, $1\leq k
\leq j$. In other words, the withdraw amount $\gamma$ takes $j$
possible values: $\gamma=A_j-A_k$, $k=1,2,\ldots,j$.

Note the above restriction that $\gamma=A_j-A_k$, $k=1,2,\ldots,j$ is
not necessary. The only real restriction is $\gamma \leq A_j$.
However, without the restriction the value of $A_j^+$ after the jump
falls between the grid points (not exactly on a grid point $A_k$)
and a costly two-dimensional interpolation is required. The error
due to this discretisation restriction can be easily reduced to
acceptable level by increasing $J$.

 For any $W=W_m$, $ m=0,1,\ldots, M$ and $A=A_j$, $
j=1,\ldots, J$ , given that withdrawal amount can only take the
pre-defined values $\gamma=A_j-A_k$, $k=1,2,\ldots,j$, irrespective
of time $t_n$ and account value $W_m$, the jump condition
(\ref{eqn_jump}) takes the following form for the specific numerical
setting

\begin{equation}\label{eqn_jump2}
Q_{t_n^-}(W_m,A_j)=\max_{1\leq k \leq j}
[Q_{t_n^+}(\max(W_m-A_j+A_k,0), A_k)+C(A_j-A_k)].
\end{equation}
For optimal strategy, we chose a value for  $1 \leq  k \leq j $ to
maximize the function value $Q_{t_n^-}(W_m,A_j)$ in
(\ref{eqn_jump2}). Note that although the jump amount
$\gamma=A_j-A_k$, $k=1,2,\ldots,j$ is independent of time $t_n$ and
account value $W_m$, the value $Q_{t_n^+}(\max(W_m-A_j+A_k,0), A_k)$
depends on all variables $(W_m,A_j ,t_n)$ and the jump amount. Thus
the above jump has to be performed for every node point $(W_m,A_j)$,
$1\leq m \leq M$, $1\leq j \leq J$ at every withdrawal date.
Obviously for every node point $(W_m,A_j)$ we have to attempt $j$
jumps to find the maximum value for $Q_{t_n^-}(W_m,A_j)$.

When $W_m-A_j+A_k > 0$, the value $Q_{t_n^+}(W_m-A_j+A_k, A_k)$ can
be obtained by interpolation from the values at the $M$ discrete
grid points. Overall we have $J$  numerical solutions (obtained
through integration)  to track, corresponding to each of the $A_j$
value, $1\leq j \leq J$. Figure \ref{fig1}  illustrates the
application of the  jump condition.

In (\ref{eqn_jump2}), to obtain $Q_{t_n^+}(W_m-A_j+A_k, A_k)$ from
solution $Q_{t_n^+}(W, A_k)$, only a one-dimensional interpolation
is required, since the coordinate in the guarantee balance space $A$
remain the same at $A_k$. Essentially we have given $M+1$ values at
$W_0,W_1,\ldots,W_M$ to find the value at $W=W_m-A_j+A_k$. For this
purpose we chose the same  cubic spline interpolation as used in
approximating the continuous function $Q_t(W,\cdot)$ in Section
\ref{sec_GHQC}.  As shown in a convergence study by
\citet{Forsyth2002}, it is possible for a PDE based numerical
algorithm for discretely sampled path-dependent option pricing to be
non-convergent (or convergent to an incorrect answer) if the
interpolation scheme is selected inappropriately. All the previous
studies of numerical PDE solution for path dependent (Asian or
lookback options) used either a linear or a quadratic interpolation
in applying the jump conditions. In our experience a better choice
is the cubic spline interpolation (\citet{Pres92}).

~

\noindent{\bf Remarks} \emph{It is worth pointing out that part of
the good efficiency of the present algorithm for pricing GMWB under
rational policyholder behavior is due to the fact that the same
cubic spline interpolation is used for both numerical integration
(\ref{eq_intS}) and the application of jump condition
(\ref{eqn_jump2}). A clear advantage of the present numerical
algorithm over PDE based finite difference approach is that
significantly smaller number of time steps are required by the
present method. In fact the number of time steps needed by the
proposed method is the same as the number of withdrawal dates, i.e.
there is no need to sub-divide the time period between  two
consecutive withdrawal dates into finer time steps - a single step
is sufficient because the transition density over the finite time
period in  (\ref{eq_intS}) is exact and there is no approximation
error due to finite time steps. On the other hand,  in general the
finite difference method requires dividing the period between two
consecutive withdrawal dates into finer time steps for a good
accuracy due to the finite difference approximation to the partial
derivatives. The above comment also applies to other derivatives
such as the American option with discrete exercise dates, Asian
options and Target Accumulation Redemption Notes, etc. with discrete
payment dates.     The accuracy of a central difference finite
difference scheme is second order both in time and space. Here the
error due to finite number of grid points in $W$ space is from cubic
spline interpolation and the Gaussian-Hermite quadrature, while in
finite difference method the error is from  the finite difference
approximation to the space derivatives. Both errors can be reduced
by increasing the number of grid points in $W$ space (reducing grid
size).}

\subsection{An alternative - GHQC with moment matching}
In calculation of annuity price expectation (\ref{eq_intY}), the probability density for $Y(t_n)$ is known in closed form. In
general the closed form pdf may not be known, and here we propose a
moment matching to replace (\ref{eq_intY}), i.e. assuming we do not
know the density in closed form but we know the moments of the
distribution, we can still use the GHQC algorithm by matching the
numerically integrated moments with the known moments. Let $p(y)$
denote the unknown density of $Y(t_n)$, then  the integration in
(\ref{eq_intY})  becomes

\begin{equation}\label{eq_intP}
Q_{t_{n-1}^+}(W(t_{n-1}),A)= e^{-r_n dt_n}\int_{-\infty}^{+\infty}p(y) Q_{t_n^-}^{(y)}(y,A) dy,
\end{equation}
which can be re-written as
\begin{equation}\label{eq_intP2}
Q_{t_{n-1}^+}(W(t_{n-1}),A)= e^{-r_n dt_n}\int_{-\infty}^{+\infty}e^{-y^2}\times[e^{y^2}p(y) ]Q^{(y)}_{t_n^-} (y,A) dy.
\end{equation}
Applying  Gauss-Hermite quadrature (\ref{eq_GHQ}) to
(\ref{eq_intP2}) we then have
 \begin{equation}\label{eq_qG}
Q_{t_{n-1}^+}(W(t_{n-1}),A)\approx    \sum_{i=1}^q \lambda_i^{(q)} \widetilde{p}(\xi_i^{(q)})
Q^{(y)}_{t_n^-}(\xi_i^{(q)}, A),
\end{equation}
where the function $\widetilde{p}(y)=e^{y^2}p(y)$
which is also unknown. Defining a new weight
$\omega_i^{(q)}=\lambda_i^{(q)} \widetilde{p}(\xi_i^{(q)})$, the numerical
quadrature for the integration simplifies to
  \begin{equation}\label{eq_qGW}
\int_{-\infty}^{+\infty}p(y) Q^{(y)}_{t_n^-}(y,A) dy \approx  \sum_{i=1}^q
\omega_i^{(q)} Q^{(y)}_{t_n^-}(\xi_i^{(q)},A).
\end{equation}
Now we proceed to find the unknown coefficients
$\omega_i^{(q)},\;i=1,2,\ldots,n$ by matching moments. Recognizing that
if we replace $Q^{(y)}_{t_n^-}(y,A)$ by $y^K$, the integration yields the $K$-th
moment corresponding to the pdf $p(y)$, thus
 \begin{equation}\label{eq_qM}
\mathrm{E}_{t_{n-1}}[Y(t_n)^K]=\int_{-\infty}^{+\infty}p(y) y^K dy =M_K(y) \approx  \sum_{i=1}^q
\omega_i^{(q)}(\xi_i^{(q)})^K,
\end{equation}
where $M_K(y)$ denotes the $K$-th moment of random variable $y$. If
we let $K=0,1,\ldots,n-1$ we then have $n$ equations to determine the
$n$ unknown coefficients $\omega_i^{(q)},\;i=1,2,\ldots,q$.

In our GMWB evaluation framework the annuity value is a function of
$X(t_n)=\ln(W(t_n)/W(0))$, and for each node point $X_m$ we have  $X(t_n)=\tau_n Y(t_n)
+\nu_n+X_m$. To
match the central moment for random variable $X(t_n)$ (centered at
$\nu_n+X_m$), equation (\ref{eq_qM}) becomes
 \begin{eqnarray}\label{eq_qMX}
\mathrm{E}_{t_{n-1}}[(X(t_n)-\nu_n-X_m)^K]&=&\int_{-\infty}^{+\infty}p_{X(t_n)}(x) (x-\nu_n-X_m)^K dx\nonumber\\
&\approx&  \sum_{i=1}^q \omega_i^{(q)}(\tau\xi_i^{(q)})^K,
\;\;K=0,1,\ldots,q-1,
\end{eqnarray}
where $p_{X(t_n)}(x)$ is the pdf for random variable $X(t_n)$.
For the standard stock market process, the central moments for $X(t_n)$
are simply
$$ \mathrm{E}_{t_{n-1}}[(X(t_n)-\nu_n-X_m)^K]=
 \left\{\begin{array}{ll}
                    0\;,\;\; & \text{if } \;K \;\text{ is odd}, \\
                   \tau^K(K-1)!!\;,\;\;&\text{if } \;K \;\text{ is even}, \\
                 \end{array}\right.    $$
where $(K-1)!!$ is the double  factorial, that is, the product of
every odd number from $K-1$ to 1.

~

\noindent{\bf Remarks}\emph{ Although in (\ref{eq_qMX}) the
Gauss-Hermite weights do not appear explicitly, it is still a direct
application of the full Gauss-Hermite quadrature. To make this
clear, we can substitute back $\omega_i^{(q)}=\lambda_i^{(q)}
\widetilde{p}(\xi_i^{(q)}) $ in (\ref{eq_qMX}) to obtain a system of linear
equations for the unknown function values
$\widetilde{p}(\xi_i^{(q)}),\;i=0,1,\ldots,q-1$
 \begin{equation}\label{eq_qMX2}
\mathrm{E}_{t_{n-1}}[(X(t_n)-\nu_n-X_m)^K] \approx  \sum_{i=1}^q\lambda_i^{(q)}
\widetilde{p}(\xi_i^{(q)})(\tau\xi_i^{(q)})^K, \;\;K=1,\ldots,q,
\end{equation}
and obviously solving (\ref{eq_qMX2}) is equivalent to solving
(\ref{eq_qMX}).}

The direct application of the Gauss-Hermite quadrature is most
suitable for the standard normal distribution, that is
 why the derivation of (\ref{eq_qMX}) is through the variable $y$. In general if the distribution of $X$ is not
 a normal distribution, we can still define $y$ through the mean and standard deviation for conditional probability of
the underlying (conditional on the value given at time step $i-1$).

Having found the $n$ coefficients $ \omega_i^{(q)}$ by solving the
system of linear equations (\ref{eq_qMX}), the expected option value
$Q_{t_{n-1}^+}\left(W(t_{n-1}^+), A(t_{n-1}^+)\right)$ is then
approximated as
\begin{equation}\label{eq_qXm}
Q_{t_{n-1}^+}\left(W(t_{i-1}^+), A(t_{i-1}^+)\right) \approx
e^{-r_n dt_n} \sum_{i=1}^q \omega_i^{(q)} Q^{x}_{t_n^-}(\tau_n
\xi_i^{(q)}+\nu_n+X_m).
\end{equation}

The GHQC algorithm with moment matching is exactly the same as the
one described earlier,  except now we have (\ref{eq_qXm}) instead of
(\ref{eq_qX}) for the numerical quadrature, and an extra step to
solve  the system of linear equations (\ref{eq_qMX}) for the new
weights to match moments. For convenience we denote the above moment
matching algorithm as GHQC-M.

\section{Numerical  Results}\label{NumericalResults_sec}
Below we present numerical results for pricing GMWB with static and optimal policyholder
strategies and compare with Monte Carlo and finite difference methods when appropriate.

\subsection{GMWB pricing results}
In general the price for a variable GMWB annuity is a function of
$(\alpha, r,\sigma,\beta, g,W(0), N_w)$.  In practice, the policyholder is charged exactly the amount of the initial investment
$W(0)$. In other words, there is no additional charge on the policyholder. This is possible only because the issuer can set the
``right" amount of the fee $\alpha$ (which is ``continuously'' taken
from the investment account), so that the annuity price $V$ equals
to the initial investment $W(0)$. Thus the pricing problem becomes:
giving $(r,\sigma,\beta, g,W(0), N_w)$, finding the correct fee
$\alpha$ so that  the annuity price $V=W(0)$. Obviously this is an
iterative process: one starts with an initial guess for the fee and
compute the annuity price, and repeats the pricing a few times while
iteratively adjust the fee value.

If the withdraw amount at each withdrawal date  is predetermined at
the beginning of the contract, then the behavior of the
 policyholder is  called ``static''. In this case the paths of account $W$ can be simulated  and a standard Monte Carlo
 simulation method can be used to price the GMWB. On the other hand if the policyholder optimally decide the amount of withdraw
 at each withdrawal date, then   the behavior of the policyholder is  called ``dynamic''.  Below we show results for both static and
  dynamic cases.  The static case allows a comparison between Monte Carlo and GHQC, further validating the new algorithm.

We  have also implemented an efficient finite difference (FD)
algorithm for pricing variable GMWB both with static and dynamic
policyholder
 behaviors.  In what follows results from GHQC
 will be compared with those from FD. In all the literature reviewed
 in the Introduction, only \citet{dai2008guaranteed} and  \citet{Forsyth2008} have presented some results for the price or fair fee of GMWB
 under the dynamic (optimal) policyholder behavior, both studies
 have  used a finite difference method. We will also compare GHQC
 results with their FD results.

\subsubsection{GMWB fair fees  with static policyholder behavior}
In a static case the withdrawal amount is pre-determined for each withdrawal date. In this case there is no
need for GHQC or FD to track many solutions for multiple levels of
the guarantee account $A$ - only one solution is required and at
each payment date  the jump condition applies to the single solution
(therefore no need for a grid in guarantee account $A$). Since the
withdrawal amount is known at every payment date, the stochastic
paths of the underlying $W$ can be simulated by Monte Carlo. Table
\ref{tab_feeS}   shows  results of the fair fee for the static case
calculated using GHQC, GHQC-M, FD and MC.
   The  withdrawal frequency is quarterly (four times per year), i.e. $N_w=4$.   The interest rate is $r=5\%$ and volatility
 is $\sigma =20\%$.  The unit of the fees (continuous rate) shown in all the following Tables  is in  basis point (bp)
 which  is $0.01\%$, i.e. a 100 basis points is $1\%$.

 For  GHQC, GHQC-M and FD the number of grids for $W$ is set at  $M=400$ . The number of time steps for FD is set at 100
 per year. The number of quadrature points for GHQC and GHQC-M is set at $q=9$.
 Unless otherwise stated, the above numerical inputs were used in all the following examples in this paper.
   We have also used $q=16$ for the number of quadrature points and found the results in the fair fee are identical at
   least in the first four digits. We observe  that for two values  of the fair fee from different methods to be identical
    in the first four digits, it requires the values of GMWB price with the same inputs to be identical in the first 6 digits.
In the Monte Carlo calculations we used  $2\times 10^7$ simulated paths (including antithetic paths).

In Table \ref{tab_feeS} the numbers
 in the parentheses (the last column) are the estimated ``standard
 errors" of the MC estimate for the fair fee. Note that the fair fee cannot be directly simulated by MC, it is inversely
 calculated in an iterative process as described earlier, thus the
 standard error of the fair fee cannot be directly estimated from MC
 samples of prices given a fee. Here we estimated the standard error
  by the difference in fees due to the standard errors in the price by the following procedure.

 Let $\widehat{Q}$ and $\epsilon_Q$ be the MC estimate of GMWB price and its  standard
  error respectively. The fair fee estimator $\widehat{\alpha}^\ast$ is obtained through $\widehat{Q}(\widehat{\alpha}^\ast)=W(0)$.
  An upper bound for the fair fee $\widehat{\alpha}^\ast_U$  can be estimated from
  $\widehat{Q}(\widehat{\alpha}^\ast_U)+\epsilon_Q=W(0)$, and a lower
  bound $\widehat{\alpha}^\ast_L$  can be estimated from
  $\widehat{Q}(\widehat{\alpha}^\ast_L)-\epsilon_Q=W(0)$. Having obtained
  the lower and upper bounds for the fair fee corresponding to the standard error in
  the price, we then estimate the standard error of the fair fee by $\epsilon_\alpha = (\widehat{\alpha}^\ast_U -
  \widehat{\alpha}^\ast_L)/2$. One can also
  took an alternative approach: numerically calculate the derivative
  $\frac{\partial \alpha}{\partial Q}$ and estimate the error in the
  fair fee by $\epsilon_\alpha =|\frac{\partial \alpha}{\partial Q}|
  \times \epsilon_Q$. We found the two approaches give us virtually
  the same answers. For example, at $g=10\%$, the first approach gives
  $\epsilon_\alpha =0.155$, while the second method yields $\epsilon_\alpha
  =0.154$. In calculating $\frac{\partial \alpha}{\partial Q}$, we
  perturb the fair fee by $1\%$ on each side of the MC estimated fair fee value and compute the corresponding price changes.

 \begin{table}[htbpc]\begin{center}
{\footnotesize{\begin{tabular*}{1.0\textwidth}{cccccc} \toprule
contractual rate, $g$ & maturity $T=1/g$ &  GHQC, bp &   GHQC-M, bp &  FD, bp  &   MC, bp \\
 \midrule
 $4\%$ & 25.00 & 17.69  &  17.69  & 17.79  &17.23 (0.120)\\
$5\%$ & 20.00 &  28.33 &  28.33  & 28.30&28.29 (0.125)\\
$6\%$ & 16.67 & 40.33  &  40.33  & 40.31 &40.37 (0.130)\\
$7\%$ & 14.29 & 53.31  &  53.31 &  53.28 &53.20 (0.135)\\
$8\%$ & 12.50 & 66.99  & 66.99 &  66.93 &67.02 (0.145)\\
$9\%$ & 11.11 & 81.23  & 81.23 & 81.21 &81.23  (0.145)\\
$10\%$ & 10.00 & 95.81  & 95.81 & 95.78&95.79 (0.155)\\
$15\%$ & 6.67 & 171.9 & 171.9  &  171.8 &171.5 (0.185)\\
\bottomrule
\end{tabular*}
}}\end{center} \caption{Fair fee $\alpha$ in bp (1 bp=0.01\%) as a function of annual
contractual rate $g$ for the static case. The parameters for the
annuity product are $r=5\%$, $\sigma =20\%$,  $N_w=4$ (quarterly
withdrawal frequency). } \label{tab_feeS}
\end{table}

 As shown in Table \ref{tab_feeS}, the fair fee is an increasing function of the contractual withdrawal rate. The GHQC and
 GHQC-M produced identical results for all the  withdrawal rates, at least for all the first 4 digits shown. Between GHQC
 and FD, the maximum absolute difference in the fee is $0.1$ bp which occurs at the highest contractual rate
 $g=15\%$. A difference of $0.1$ basis point in the fee $\alpha$ is about 1 cent per year for a $\$1000$ account.
  Between GHQC and MC, the maximum absolute difference in the fee is $1.2$ bp,  about 12 cents a year for a
  $\$1000$ account,  which also occurs at the highest contractual rate $g=15\%$.
 The computing time requirement for both FD and HGQC in the static case is very fast - both took a fraction of a second to compute a single price. We will have a detailed comparison of the
   computing speed in the next section dealing with pricing GMWB in the dynamic case, where the computation is much more
    demanding because multiple solutions of many
   levels of guarantee amount have to be tracked and multiple jumps have to be applied for finding the optimal strategy.

\subsubsection{GMWB fair fees with optimal policyholder behavior}

Table \ref{tab_feeQ}   shows  results of the fair fee for the
dynamic case at a quarterly withdrawal frequency ($N_w=4$),
calculated using GHQC and FD. In this example the number of grids in
the guarantee account $A$ is $J=100$ for both GHQC and FD.  The
maximum difference between GHQC and FD in the calculated fair fee is
$0.15$ bp,  less than  2 cents a
 year for a $\$1000$ account,  which occurs at the lowest contractual rate $g=4\%$. Again the
  GHQC-M results (not shown in the table) were identical to those from GHQC in at least the first four digits shown.

\begin{table}[htbpc]
\begin{center}
{\footnotesize{\begin{tabular*}{0.65\textwidth}{cccc} \toprule
contractual rate, $g$ & maturity $T=1/g$ &  GHQC, bp &   FD, bp \\
 \midrule
 $4\%$ & 25.00 & 56.09 & 55.94  \\
$5\%$ & 20.00 &  70.06 &  69.96 \\
$6\%$ & 16.67 &  83.73  &  83.64 \\
$7\%$ & 14.29 & 97.11 &  97.03 \\
$8\%$ & 12.50 & 110.3  &110.2  \\
$9\%$ & 11.11 & 123.2& 123.1  \\
$10\%$ & 10.00 & 136.0 & 135.9  \\
$15\%$ & 6.67 & 199.0 & 199.0 \\
\bottomrule
\end{tabular*}
}} \end{center}\caption{Fair fee $\alpha$ in bp (1 bp=0.01\%) as a function of annual
contractual rate $g$ for the dynamic case with a quarterly withdrawal
frequency ($N_w=4$). The other parameters for the GMWB product are
$r=5\%$, $\sigma =20\%$, $\beta=10\%$.} \label{tab_feeQ}
\end{table}

Compared with the static case, the fees for the dynamic case are
much higher - the dynamic fee is $15\%$ higher than the
 static fee  at the contract rate $g=15\%$ and this difference increases to $217\%$ at  $g=4\%$. Recall
 that the maturity $T=1/g$, so  the lowest rate corresponds to the longest maturity and vise versa. With a fixed
 withdrawal frequency, a longer maturity means more opportunities for the policyholder to make optimal decisions to
 maximize the total cashflow from holding the variable GMWB contract. Thus at the lowest contract  rate the fee for the
 dynamic case shows the highest percentage increase from the static case, highlighting the value of optimal decisions
 under uncertainty.

  \begin{table}[htbpc]
  \begin{center}
{\footnotesize{\begin{tabular*}{0.9\textwidth}{ccccc} \toprule
contractual rate, $g$ & maturity $T=1/g$ &  GHQC, bp &   FD, bp  & \citet{dai2008guaranteed}, bp \\
 \midrule
 $4\%$ & 25.00 & 56.77  &56.68  & 56  \\
$5\%$ & 20.00 & 70.92  &  70.78 & 69 \\
$6\%$ & 16.67 & 84.76&  84.63  &83 \\
$7\%$ & 14.29 & 98.30 &  98.18  & 97 \\
$8\%$ & 12.50 & 111.6  &111.5 &  111  \\
$9\%$ & 11.11 & 124.7 & 124.6  & 124  \\
$10\%$ & 10.00 & 137.7  & 137.5 & 137  \\
$15\%$ & 6.67 & 201.7 & 201.6 & 198  \\
\bottomrule
\end{tabular*}
}} \end{center}\caption{Fair fee $\alpha$ in bp (1 bp=0.01\%) as a function of annual
contractual rate $g$ for the dynamic case with a monthly withdrawal
frequency ($N_w=12$). The other parameters for the GMWB product are
$r=5\%$, $\sigma =20\%$, $\beta=10\%$.} \label{tab_feeM}
\end{table}

  Table \ref{tab_feeM}   shows  results of the fair fee for the dynamic  case at a monthly withdrawal frequency ($N_w=12$)
  calculated using GHQC and FD.  For these monthly withdrawal cases the number of grids in $A$ was set at $A=300$ for both
  GHQC and FD.  The maximum  difference  between GHQC and FD in the calculated fair fees shown in  Table \ref{tab_feeM}
  is $0.2$ basis point,  about 2 cents a year for a $\$1000$ account.
  Comparing Table  \ref{tab_feeM} and Table  \ref{tab_feeQ},  our results from both GHQC and FD consistently show a
  higher fee for  a higher withdraw frequency at all contractual rates. In general a higher withdrawal frequency  should
  have a higher
 fair fee than a lower withdrawal frequency, since the
 former  has a higher annuity value for the same fee. The higher withdrawal frequency allows the policyholder to have
  more opportunities to make optimal decisions to maximize the total cashflow and so it is more valuable.
   Nevertheless the relative difference in fees  between monthly and quarterly withdrawal contracts is in the order
    of $1\%$ only. In absolute terms the maximum difference in the fees  between the monthly and quarterly withdrawal
    contracts is less than 3 basis points, occurring at the
  highest contractual rate $g=15\%$. A difference of 3 basis points in fees is only about $30$ cents per year for a
   $\$1000$ account value. The close agreement in fees (or prices) between the quarterly and monthly withdrawal
   frequency indicates that
   the results for the monthly  withdrawal frequency should already be approaching  those of the continuous case.

  Also shown in Table \ref{tab_feeM} are the results for a continuous withdrawal model obtained by \citet{dai2008guaranteed} solving
   a two dimensional linear complementary problem using a penalty finite difference method.
    In the case of a continuous withdrawal model,  the policyholder can make a decision at any instance of
time and the withdraw amount
 can be infinitely small or finite, thus it can be more optimal than
 any discrete case.
 However, our results for the discrete withdrawal model at a monthly frequency are slightly higher than those of
  \citet{dai2008guaranteed} for the continuous case, which should not happen if  the numerical calculations for both the
  discrete model and  the continuous model are  exact. Still, as can be calculated from  Table \ref{tab_feeM}, the
  difference in the fees between our monthly withdrawal model and the continuous model is small - the maximum absolute
  difference is $3.7$ basis points occurring at the highest contractual rate
  $g=15\%$.

 \begin{table}[htbpc]\begin{center}
{\footnotesize{\begin{tabular*}{0.6\textwidth}{cccc} \toprule
contractual rate, g & maturity T=1/g &  GHQC &   FD   \\
 \midrule
 $4\%$ & 25.00 & 102.0  &101.3   \\
$5\%$ & 20.00 & 123.6  &  123.2 \\
$6\%$ & 16.67 & 144.1 &  143.7  \\
$7\%$ & 14.29 & 163.5&  163.2   \\
$8\%$ & 12.50 & 182.1  &181.8   \\
$9\%$ & 11.11 & 199.8 & 199.6  \\
$10\%$ & 10.00 & 216.9  & 216.7  \\
$15\%$ & 6.67 & 293.8 & 293.8 \\
 \bottomrule
\end{tabular*}
}}\end{center} \caption{Fair fee $\alpha$ as a function of annual
guarantee rate $g$ for the dynamic case with a quarterly withdrawal
frequency ($N_w=4$) and with the penalty rate reduced to
$\beta=5\%$. The other parameters for the GMWB product are $r=5\%$,
$\sigma =20\%$.} \label{tab_feeQ2}
\end{table}

\begin{figure}[htbp]
\begin{center}
\includegraphics[scale=0.9]{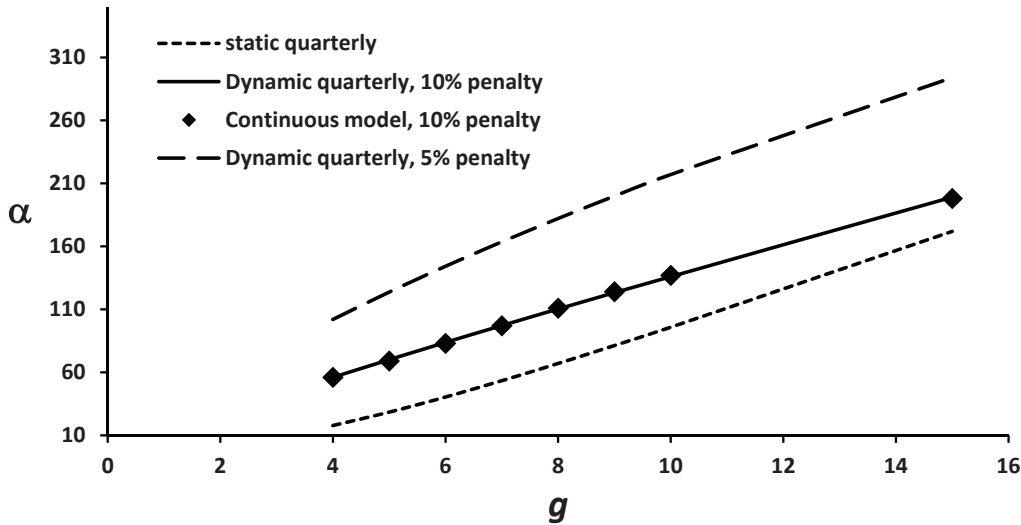}
\vspace{0cm} \caption{Fair fee $\alpha$ as a function of annual
guarantee rate $g$ for three discrete withdrawal contracts at a
quarterly withdraw rate, in comparison with the continuous
model.}\label{fig_fee}
\end{center}
\end{figure}

  Table \ref{tab_feeQ2} shows results for a quarterly withdrawal contract when the penalty rate $\beta$ is reduced
  to $5\%$. As a result of the  reduced penalty rate, the fair fees increase very significantly: at $g=15\%$ (shortest
   maturity) the fee is increased by $45\%$ and this increase  becomes almost $80\%$ at $g=4\%$ (the longest maturity).
   This demonstrates that  the optimal decision is much more valuable when it is not restricted by a penalty.
    To illustrate the relative magnitudes of the fees for various cases, Figure \ref{fig_fee} compares the fair
    fees for  four cases: static quarterly withdrawal, dynamic quarterly withdrawal with $10\%$ penalty charge,
    dynamic quarterly withdrawal with $5\%$ penalty charge,  and the continuous withdrawal model  calculated by
    \citet{dai2008guaranteed}.  In   Figure \ref{fig_fee} data for all the three curves of the  discrete withdrawal model are
    calculated using the GHQC algorithm. Results of FD and GHQC-M for the same cases are not shown because they appear
    identical on the graph.

To compare computing speed between FD and GHQC, we record the CPU
time for a single price calculation - the calculation of the fair
fee involves many such calculations in an iterative process. The CPU
used for all the calculations in this study is
 Intel(R)  Core(TM) i5-2400 @3.1GHz.
 Apart
from the numerical mesh size and time steps, the CPU time for a
single price calculation for a dynamic case also depends on the
maturity and withdrawal frequency. Longer maturity and higher
frequency demand more computing time. Among  the examples shown in
Table \ref{tab_feeQ} and   Table  \ref{tab_feeM}, the case with a
quarterly withdraw frequency at the shorted maturity ($g=15\%$ and
$T=1/g=6.67$ years) is the least demanding in computing time. For
this case the FD took 14 seconds and the GHQC took 2 seconds to
compute a single price. On the other end, the case with a monthly
withdraw frequency at the longest maturity ($g=4\%$ and $T=1/g=25$
years) is the most demanding in computing time, and  for this case
the FD took 482 seconds and the GHQC took 167 seconds to compute a
single price. As discussed earlier,  both FD and GHQC used the same
grids for both $A$ and $W$.  Obviously the speed advantage of GHQC
is more pronounced at lower withdrawal frequency - only a single
time step is required for GHQC between consecutive withdrawal dates,
while for FD the time step size has to be sufficiently small for
good accuracy, irrespective of the withdrawal frequency. For
example, to price a GMWB at a yearly withdrawal frequency, the GHQC
is more than 15 times as fast as FD using the same grids in $W$ and
$A$. Note that we believe our finite difference implementation for
pricing GMWB is already very efficient - for example the
tri-diagonal matrix for the linear equation discretising the PDE  is
constructed and inverted only once for all the constant time steps
and for all the solutions at all levels of the guarantee amount $A$,
taking advantage of the constant interest rate and volatility. So
each time step in the FD only involves a simple back-substitution
which takes little CPU time.

\begin{table}[htbpc]
\begin{center}
{\footnotesize{\begin{tabular*}{0.6\textwidth}{cccc} \toprule
 frequency & volatility &   \citet{Forsyth2008} &   GHQC   \\
 \midrule
 yearly      & 0.2 & 129.1 & 129.1   \\
 half-yearly & 0.2 & 133.5 & 133.7 \\
 yearly      & 0.3 & 293.3 & 293.5  \\
 half-yearly & 0.3 & 302.4 & 302.7   \\
\bottomrule
\end{tabular*}
}}\end{center} \caption{Comparison of fair fee $\alpha$ between
results of GHQC and those from finite difference by Chen and Forsyth
(2008). The input parameters are $g=10\%$, $\beta=10\%$, $r=5\%$.}
\label{tab_g10}
\end{table}

In  \citet{Forsyth2008}, the fair fees for the
  discrete withdrawal model with $g=10\%$ for the yearly ($Nw=1$) and
  half-yearly ($Nw=2$) withdrawal frequency at $\sigma=0.2$ and $\sigma=0.3$  were presented in a carefully performed
  convergence study, with the same values for other input
  parameters ($r=5\%$, $\beta=10\%$). Table \ref{tab_g10} compares
  GHQC results with those of  \citet{Forsyth2008}. The values of
   \citet{Forsyth2008} quoted in Table \ref{tab_g10} correspond to their finest mesh grids and time
  steps at $M=2049$ for $W$, $J=1601$ for $A$ and $N=1920$ for $t$, while  our GHQC values were
  obtained using $M=400$, $J=100$ and with $n=9$ for the number of quadrature
  points. As shown in Table \ref{tab_g10}, the maximum absolute
  difference in the fair fee rate between the two numerical studies is only $0.3$ bp, and the average absolute
   difference of the four cases in the table is less than 0.2 bp. In relative terms, the maximum difference is less than $0.15\%$, and the average magnitude of the relative differences between
   the two studies is less than $0.08\%$.  \citet{Forsyth2008} did not provide CPU numbers for
   their
   calculations of fair fees. In our case each calculation of the
   fair fee in Table \ref{tab_g10}, which involves a Newton
   iteration, took about 5 seconds.

  \section{Conclusion}\label{conclusion_sec}
   In this
paper we have presented a new  algorithm for pricing variable
annuities with GMWB features under both static and dynamic (optimal)
policyholder behaviors. The new method is neither  based on solving
PDEs using finite difference nor on simulations using Monte Carlo.
The new algorithm relies on numerically  computing the expected
option values in a backward time-stepping between withdrawal dates
through Gauss-Hermite integration quadrature applied on a cubic
spline interpolation, either using transition density or matching
moments.
  At discrete withdrawing
dates the proper jump conditions are applied at various withdrawal
account levels  which allows the optimal withdrawal
 decision to be made.

Numerical results from the new algorithm for  a series of GMWB
contract are presented, in comparison with  results
 using finite difference method. The comparison  demonstrates that the new algorithm produces  results very close to those of
 the finite difference (from our own or from literature), but at the same time  it is significantly
 faster. Using the same grids for
 the space variables for the underlying  and the guarantee amount, the relative difference in price between
 GHQC and FD
  is in the order of $0.001\%$, and  the absolute
  difference in the fair fee rate between the two methods is in the order of $0.1$ bp, which
  is one cent per year in a $\$1000$ account.

So far we have assumed the policyholder will always live beyond the
maturity date or there is always someone there to make optimal
withdrawal decisions for the entire duration of the contract. In
reality an elderly  policyholder may die before maturity date,
especially for a contract with a  long maturity. For example,
according to the Australian Life Table, a male aged 60 will have
more than $57\%$ probability to die before the age of 85. So for  a
60 year old male taking up a GMWB contract with a maturity of 25
years ($g=4\%$), it is unrealistic not to consider the probability
of death during the contract period and design and pricing the
suitable contract accordingly. It is not difficult to  consider
adding some death benefits on top of GMWB, i.e. combining GMWB with
some kind of life insurance. The new method will enable us to
 efficiently explore possibilities of new products considering both market
process and death process. Further work includes admitting other
stochastic risk factors such as stochastic interest rate or
volatility. It remains to be seen whether the new GHQC algorithm can
 still be significantly faster than the finite difference method in
 higher dimensions for solving the optimal stochastic control
 problem arising from pricing GMWB under the dynamic policyholder
  behavior.


\bibliographystyle{chicago} 
\bibliography{bibliography}

\end{document}